\documentclass[%
 reprint,
 amsmath,amssymb,
 aps,
]{revtex4-2}

\usepackage{graphicx}
\usepackage{dcolumn}
\usepackage{bm}
\usepackage{aas_macros}
\usepackage{xcolor}

\begin{document}

\preprint{APS/123-QED}

\title{A common origin of multi-messenger spectral anomaly of galactic cosmic rays}

\author{Yu-Hua Yao}
\affiliation{%
 College of Physics, Chongqing University, No.55 Daxuecheng South Road, High-tech District, Chongqing, 401331, China
}%
 \affiliation{Key Laboratory of Particle Astrophysics, Institute of High Energy Physics, Chinese Academy of Sciences,
 No.19 B Yuquan Road, Shijingshan District, Beijing, 100049, Beijing, China
 }

\author{Xu-Lin Dong}%
\affiliation{College of Physics, Hebei Normal University, No.20 Road East 2nd Ring South, Shijiazhuang, 050024, Hebei, China
}

\author{Yi-Qing Guo}
\email{guoyq@ihep.ac.cn}
\affiliation{
Key Laboratory of Particle Astrophysics, Institute of High Energy Physics, Chinese Academy of Sciences,
 No.19 B Yuquan Road, Shijingshan District, Beijing, 100049, Beijing, China
}%
\affiliation{
College of Physics, University of Chinese Academy of Sciences, No.19 A Yuquan Road, Shijingshan District, Beijing, 100049, Beijing, China
}%

\author{Qiang Yuan}
\email{yuanq@pmo.ac.cn}
\affiliation{%
Key Laboratory of Dark Matter and Space Astronomy, Purple Mountain Observatory, Chinese Academy of Sciences, No.10 Yuanhua Road, Qixia District, Nanjing, 210023, Jiangsu, China
}%
\affiliation{School of Astronomy and Space Science, University of Science and Technology of China, Hefei 230026, China
}

\date{\today}

\begin{abstract}
Cosmic ray (CR) spectra show a two-component structure for both primary and secondary particles. In our study, we think that this feature roots in a nearby-source halo. This halo acts as a reservoir for CRs in the source region, leading to the high-energy component. Furthermore, it acts as a barrier, effectively halting the travel of CRs from other sources directly in the disk, this causes CRs to propagate from the outer disk and contribute to the low-energy component. As a result, all the spectra anomalies could be well reproduced. This scenario can be extended to the whole disk, we prospect high-energy diffuse $\gamma$-ray emissions exhibit highly spatial dependence, emanating from local sources. We urge HAWC/LHAASO to detect diffuse $\gamma$ ray on the spatial scale of a few square degrees.
\end{abstract}

\maketitle

\section{Introduction}

\label{introduction}
\begin{figure*}[!htb]
\centering
\includegraphics[width=0.85\textwidth]{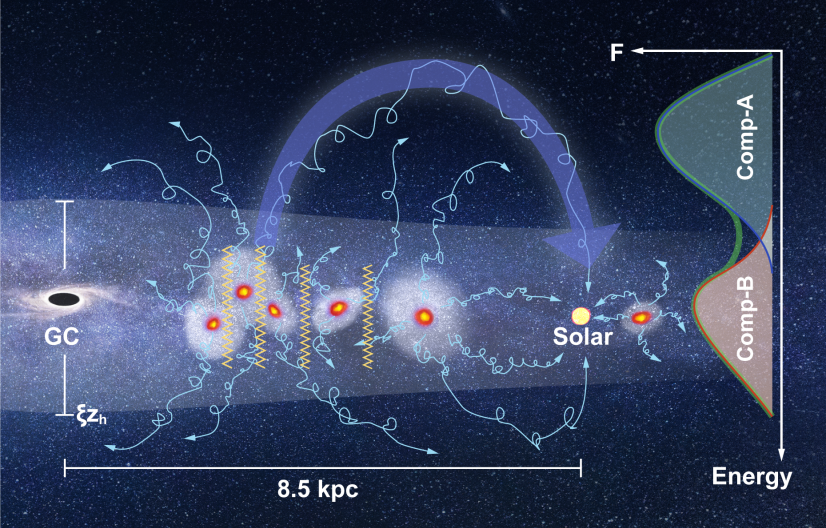}
\caption{The cartoon of the propagation of CRs in two diffuse regions and the underlying physical mechanism responsible for generating the two-component spectra. The Galactic plane is populated with halos, which serve as barriers hindering CRs from other sources from traversing this region, while also functioning as CR wells that confine CRs locally. Low-energy CRs predominantly originate from remote sources, while high-energy CRs are primarily contributed by nearby sources.
}
\label{fig:cartoon}
\end{figure*}

The latest generation of detectors has enabled precise measurements of cosmic ray (CR) spectra, revealing intriguing and unexpected features. The spectra of all nuclei exhibit a hardening at a rigidity of approximately 200 GV \citep{2009BRASP..73..564P,2011Sci...332...69A,2011ApJ...728..122Y,2019SciA....5.3793A,2018PhRvL.120b1101A,2020PhRvL.124u1102A,2021PhRvL.126d1104A,2021PhR...894....1A}. The flux ratios of $\bar{p}/p$ and B/C also display similar structures at around 20 and 200 GV, respectively \citep{2016PhRvL.117i1103A,DAMPECOLLABORATION2022}. The electron (positron) spectrum shows a hardening above 42 (10) GeV \citep{2014PhRvL.113l1102A,2019PhRvL.122d1102A,2017Natur.552...63D}. Moreover, the diffuse $\gamma$-ray emission in the Galactic disk exhibits an unexpected excess above 20 GeV \citep{2012ApJ...750....3A}. Given these unexpected anomalies, it is crucial to investigate their intrinsic relationship and shared properties.

These spectra can be classified into two categories: primary and secondary. The primary spectra exhibit a clear hardening at around 200 GV \citep{2019SciA....5.3793A,2018PhRvL.120b1101A}. The spectral characteristics of secondary particles are determined by those of their parent particles and are influenced by the interaction modes. Secondary nuclei resulting from fragmentation interactions, such as Li, Be, and B, exhibit the same spectral structure as their parent C, N, and O. Proton-proton collision interactions can generate secondary particles such as $\rm\bar{p}$, $\rm\gamma$, and $\rm e^{+}$. The energy of $\rm\bar{p}$ and $\rm\gamma$ is approximately 10\% of their main parent proton, while $\rm e^{+}$ accounts for about 5\% \citep{2023ApJ...956...75Q}. This implies that the $\bar{p}/p$ distribution and diffuse $\rm\gamma$-ray exhibit hardening at around 20 GeV, while $\rm e^{+}$ shows hardening at around 10 GeV. In short, the primary spectra and the parents of secondary particles share a two-component spectral structure at approximately 200 GV, referred to as "Comp-A" and "Comp-B" for the low-energy and high-energy components, respectively.

Investigating the origin of Comp-A and Comp-B is a natural step in understanding these spectral components. Various theoretical models have been proposed to explain the observed anomalies, primarily focusing on the reacceleration mechanism of old supernova remnants \citep{2010ApJ...725..184B,2014A&A...567A..33T}, the presence of specific sources \citep{2012MNRAS.421.1209T}, or spatially-dependent propagation (SDP) \citep{2012ApJ...752L..13T,2015PhRvD..91h3012G,2016ChPhC..40a5101J}. However, explaining the secondary spectra within the framework of the reacceleration mechanism remains a challenge. Source models incorporate a freshly accelerated component into the background CR to account for the observed spectral features, but they struggle to reproduce the diffuse $\rm \gamma$-ray emission. Additionally, previous works based on SDP models were lack of a clear description of the origin of the two spectral components.

Recently, significant progress has been achieved through the discovery of the source halo, offering fresh insights into this field of study. The Geminga halo, characterized by significantly slower dissemination, was first identified by the HAWC collaboration \citep{2017Sci...358..911A}. Subsequent findings from the LHAASO experiment further supported the presence of halos surrounding pulsars as a common occurrence \citep{2021PhRvL.126x1103A}. To date, seven halos have been measured \footnote{http://tevcat.uchicago.edu/}, sparking extensive discussions regarding the contribution of TeV halos to the overall population of TeV-emitting pulsar wind nebulae \citep{2017PhRvD..96j3016L,2022NatAs...6..199L}. It is plausible that a universal slow diffuse zone, encompassing all pulsars with slow diffuse halos, exists within the galactic plane.

In this work, we hope to present a clear physical picture for the origin of Comp-A and Comp-B. Our research investigates the contributions of different source locations to particle spectra, expanding this concept to encompass both local and remote sources across the entire Galaxy. The organization of this paper is as follows: Section 2 presents CR SDP model, followed by the results in Section 3. Finally, Section 4 provides a summary.

\section{Model $\&$ Methodology}
The newly discovered source halo sheds new perspectives to this study for the two-component structure of CR spectra. We think that this feature originates from nearby-source halo. The key point is that the halo acts as not only a CR well but also a barrier, as shown in Figure \ref{fig:cartoon}. It's a well that confines newly source-accelerated particles in this region. The halo-trapped particles exhibit an extended retention period, resulting in the maintenance of the source injection spectrum and a hardening spectral index. This part is the Comp-B, which is fresh, local, and specific in the galactic disk. Its characteristics, like the intensity, spectra index, break-off, are directly related to the halos. Simultaneously, the halo is a barrier that prevents CRs from other sources from directly traverse the disk, as a result, they are likely to undergo rapid dispersion in regions outside of the disk. The faster diffusion process in the outer disk, compared to the Galactic disk, leads to a softer spectrum and the formation of Comp-A. This component, known as the "CR sea," is a long-lasting, global, and universal feature of the Galaxy.

 In the following a quantitative description of the CR propagation process is provided. The propagation of CRs can be represented mathematically by the following partial differential equation,
\begin{eqnarray}
\frac{\partial \psi(\vec{r}, p, t)}{\partial t} &=& Q(\vec{r}, p, t) + \vec{\nabla} \cdot ( D_{xx}\vec{\nabla}\psi - \vec{V}_{c}\psi )
\nonumber\\
&& + \frac{\partial}{\partial p}\left[p^2D_{pp}\frac{\partial}{\partial p}\frac{\psi}{p^2}\right]
\nonumber\\
&& - \frac{\partial}{\partial p}\left[ \dot{p}\psi - \frac{p}{3}(\vec{\nabla}\cdot\vec{V}_c)\psi \right]
\nonumber\\
&&- \frac{\psi}{\tau_f} - \frac{\psi}{\tau_r} ~,
\label{propagation_equation}
\end{eqnarray}
where $ Q(\vec{r}, p, t)$ describes the distribution of sources, $\vec{V}_{c}$ represents the convection velocity, $ D_{xx}$ denotes the spatial diffusion coefficient, $D_{pp}$ is the momentum diffusion coefficient accounting for the re-acceleration process, $\dot{p}$, $\tau_f$, and $\tau_r$ are the energy loss rate, the fragmentation and radioactive decaying time scales, respectively.

To delineate the confinement of CRs within a cylindrical region during their propagation, we adopt the SDP model, which partitions the diffusive zone into two symmetric regions characterized by different diffusion coefficients. Specifically, the inner zone surrounding the Galactic disk exhibits slower CR dispersion compared to the outer zone. The diffusion coefficient, denoted as $ D_{xx}$, is parameterized as follows:
\begin{equation}
    D_{xx}(r, z, \mathcal R) = D_0 F(r, z) \beta^\eta \left(\frac{\mathcal R}{\mathcal R_0} \right)^{\delta_0 F(r, z)} ~,
\end{equation}
where the function $F(r,z)$ is defined as:
\begin{equation}
   F(r,z) =
   \begin{cases} g(r,z) +\left[1-g(r,z) \right] \left(\dfrac{z}{\xi z_0} \right)^{n} , &  |z| \leq \xi z_0 \\
1 ~, & |z| > \xi z_0
    \end{cases},
\end{equation}
with $g(r,z) = N_m/[1+f(r,z)]$. Here the diffusion coefficient of the inner zone is anti-correlated with the source distribution $f(r,z)$, given by
\begin{equation}
f(r, z)=\left(\frac{r}{r_{\odot}}\right)^{1.25} \exp \left[-\frac{3.87\left(r-r_{\odot}\right)}{r_{\odot}}\right] \exp \left(-\frac{|z|}{z_s}\right),
\end{equation}
where $r_{\odot}$ = 8.5~kpc and $z_s$ = 0.2~kpc. For the outer zone, the diffusion coefficient remains constant when varying spatial locations. As a comparison, for the traditional model as given in Appendix A, the diffusion coefficient keeps a constant throughout the entire space. Note that, other parameters such as the source injection parameters for the traditional model are not tuned to fit the data. This is because we do not aim to compare the goodness-of-fit of these two models, but just to illustrate the differences on interpreting the low- and high-energy spectra of the data under these two scenarios. To investigate the origin of particle spectra at different energy ranges in relation to the source locations, we divide $f(r,z)$ into two parts at distinct locations. One part corresponds to nearby sources, while the other represents remote sources. For the purpose of comparing charged particle spectra, the central location is defined as the position of the Sun.

The source spectrum of CRs is assumed to be a broken power law in rigidity. To solve the transport equation, we employ the numerical package DRAGON \citep{2008JCAP...10..018E}. The force-field approximation \cite{1968ApJ...154.1011G} is incorporated to account for solar modulation effects. The key parameters pertaining to CR propagation are summarized in Table \ref{tab:para}.


\begin{table*}[!htb]
\caption{Parameters of the SDP propagation model.}\label{tab:para}%
\begin{tabular*}{\textwidth}{@{\extracolsep\fill}ccccccc}
\hline
 $ D_0^{\dagger}  $  & $ \delta_{0}$ & $ N_m$ &  $\xi$ & n      &  $ v_{A}$       & $ z_0$ \\
 $ [cm^{-2}~s^{-1}]$ &               &        &        &        &  $ [km~s^{-1}]$ & [kpc] \\
 \hline
 $4.8 \times 10^{28}$& 0.63          &  0.24  & 0.1    & 4.0    & 6               & 4.5 \\
 \hline
 $^\dagger$Reference rigidity is 4 GV.
\end{tabular*}
\end{table*}

\section{Results}
Applying the aforementioned methodology, we computed the spectra of protons, the $\rm B/C$ and $\rm \bar p/p$ flux ratios. These calculations were then extended to cover the entire galactic plane, allowing for an investigation of the spectral index and density variations of CRs across the Galaxy.

\begin{figure}[!htb]
\centering
\includegraphics[width=0.49\textwidth]{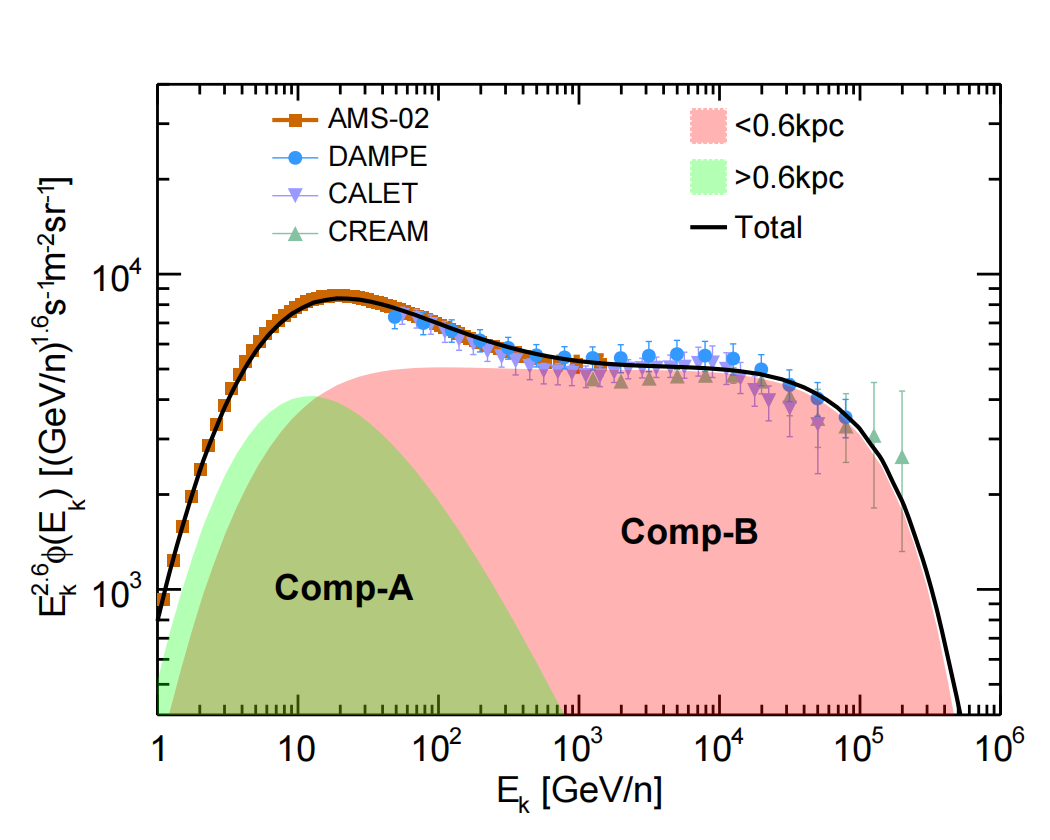}
\caption{
Proton spectrum from model calculations and observations by AMS-02 \citep{2015PhRvL.114q1103A}, DAMPE \citep{2019SciA....5.3793A}, and CREAM-III \citep{2017ApJ...839....5Y}. The pink and green shadow regions represent fluxes contributed by sources within the $ r_s<$0.6~kpc and $ r_s>$0.6~kpc regions, respectively. The black solid line represents their total fluxes. An energy cut-off of 1.4$\times10^{5}$~GV is applied for the proton flux originating from nearby sources (i.e., within $r_s <$ 0.6~kpc).
}
\label{fig:p}
\end{figure}

\begin{figure*}[!htb]
\centering
\includegraphics[width=0.49\textwidth]{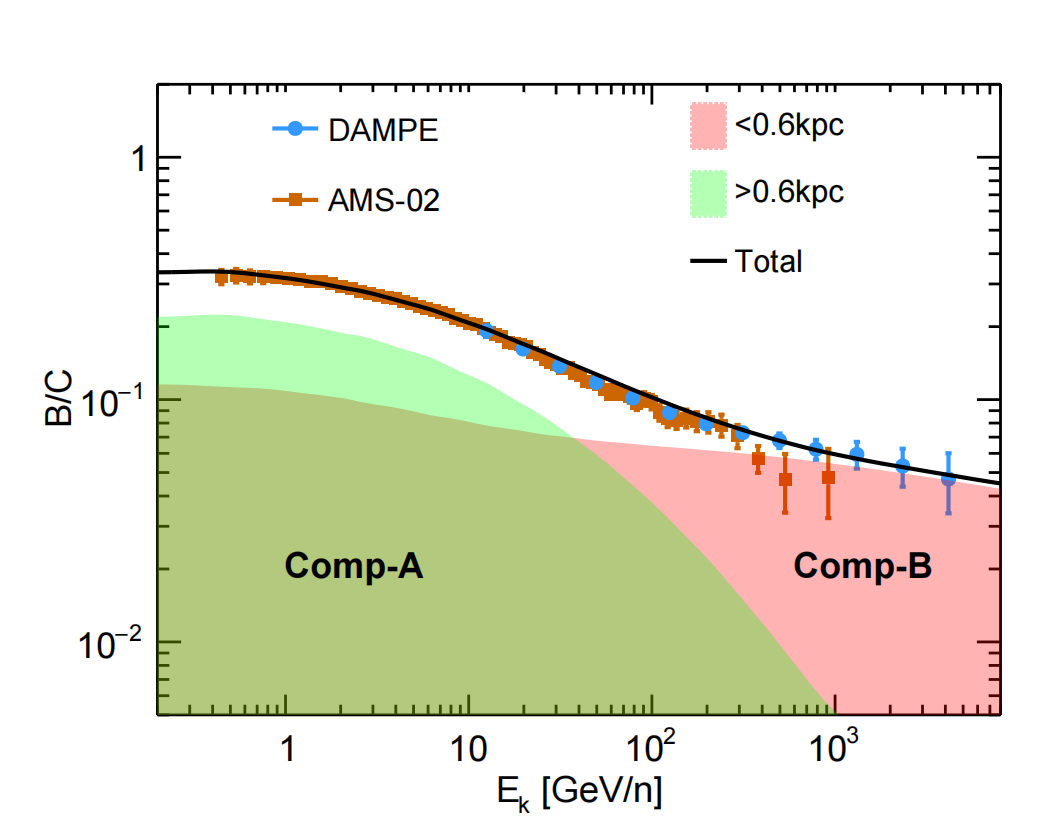}
\includegraphics[width=0.49\textwidth]{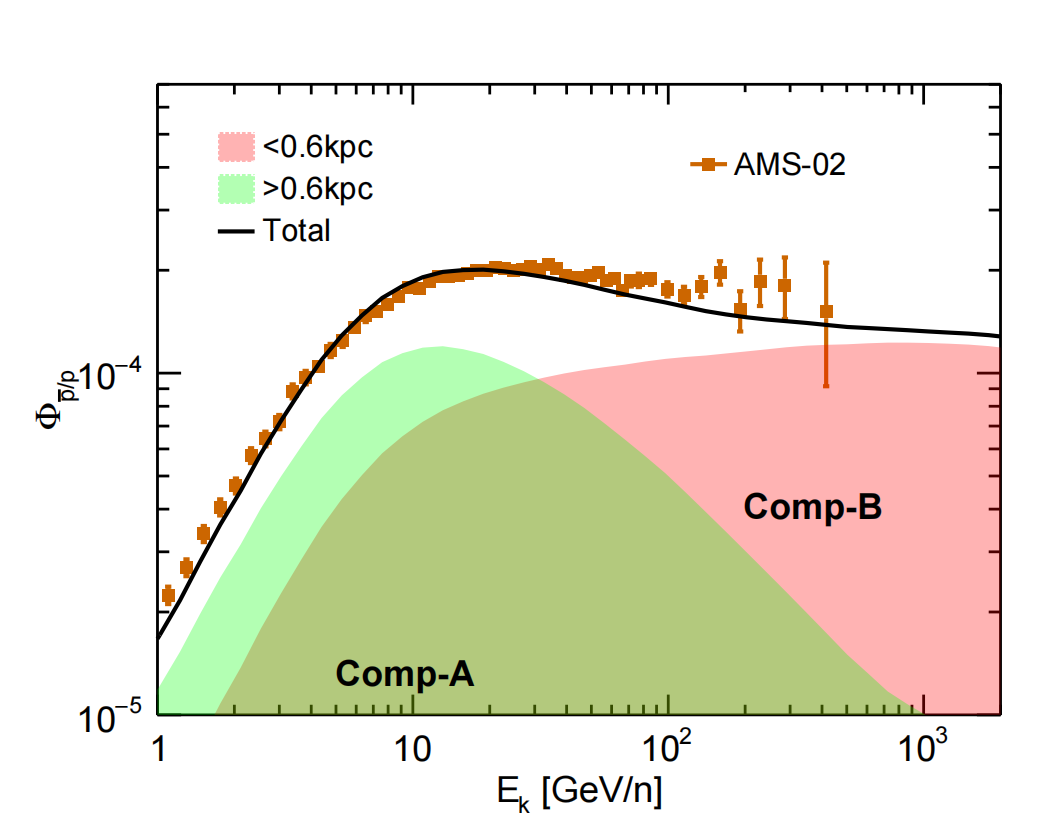}
\caption{SDP model calculated ratios of $\rm B/C$ (left) and $\rm \bar p/p$ (right) compared with measurements from AMS-02 \citep{2016PhRvL.117i1103A,2016PhRvL.117w1102A} and DAMPE \citep{DAMPECOLLABORATION2022}. The shadow regions in pink and green represent fluxes contributed by sources within the $ r_s <$ 0.6~kpc and $ r_s >$ 0.6~kpc regions, respectively. The black solid lines represent total fluxes from the entire region.
}
\label{fig:pbarp}
\end{figure*}

Figure \ref{fig:p} presents a comparison between the proton spectra obtained from observations and the predictions of the SDP model. It is evident that the calculated flux from the SDP model accurately reproduces the observed spectral features. To investigate the variations in the contribution of sources to the overall flux in different regions, the source density distribution $f(r,z)$ is divided into two zones centered at the Sun. The first zone encompasses a spherical volume with radii $r_s < $ 0.6~kpc, while the second zone includes the remaining region with radii $ r_s >$ 0.6~kpc, based on the height of the inner diffusive zone ($\xi~z_0$). Notably, the contribution from distant sources ($ r_s >$ 0.6~kpc) is Comp-A. This region of the CR spectrum is affected by the rapid diffusion occurring outside the galactic disk, and is primarily characterized by low-energy phenomena. On the other hand, the contribution from nearby sources ($r_s <$ 0.6~kpc) corresponds to Comp-B, which is confined within a slower diffusive region and has limited extension into the outer diffusive zone beyond the galactic disk. To reproduce the observed spectral break at 14 TeV \citep{2019SciA....5.3793A}, we employed an exponential cutoff model at 1.4$\times10^5$~GV to describe the proton injection spectrum for local sources within regions where the radial distance ($r_s$) is less than 0.6 kpc from the Sun.

Figure \ref{fig:pbarp} shows calculated ratios of $\rm B/C$ and $\rm\bar p/p$, comparing with the observations \citep{2016PhRvL.117i1103A,2016PhRvL.117w1102A,DAMPECOLLABORATION2022}. Comp-A has a quick fall, similar to the structure of the proton spectrum, and is dispersed throughout the galaxy as a sea of secondary particles. Comp-B is unique to the galactic disk and dominates at high energy. This result favours the general picture as Figure \ref{fig:cartoon}.

\begin{figure*}[!htb]
\centering
\includegraphics[width=0.49\textwidth]{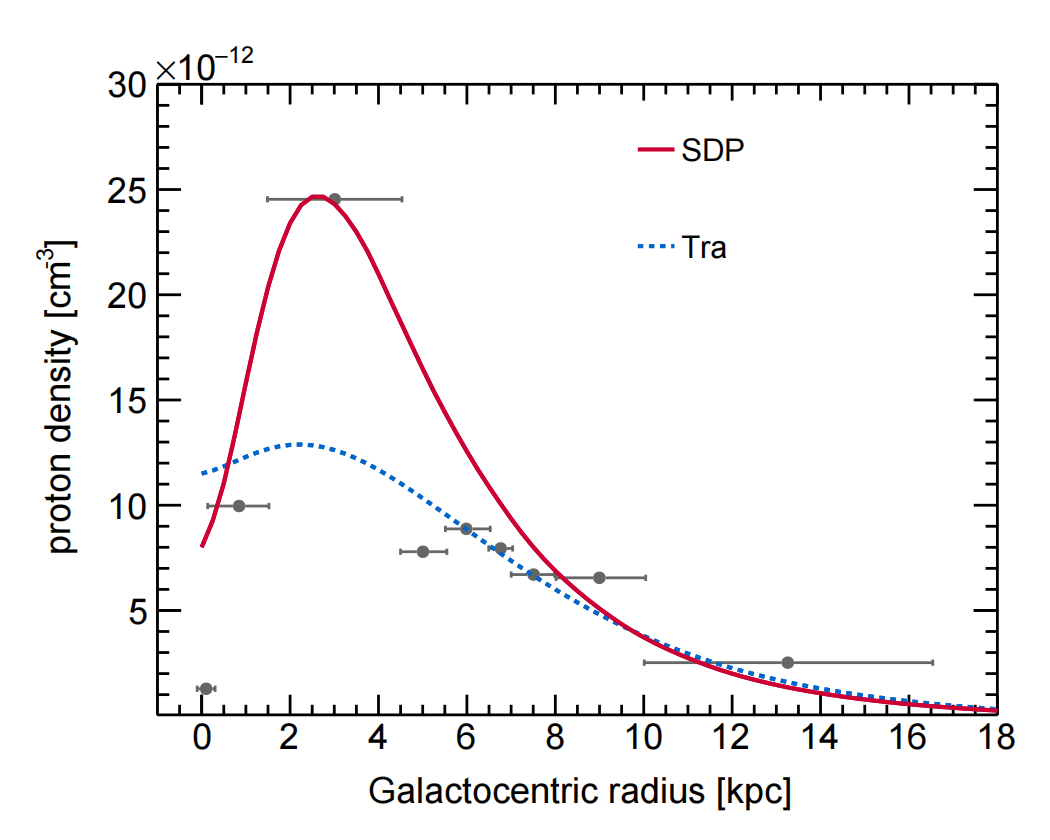}
\includegraphics[width=0.49\textwidth]{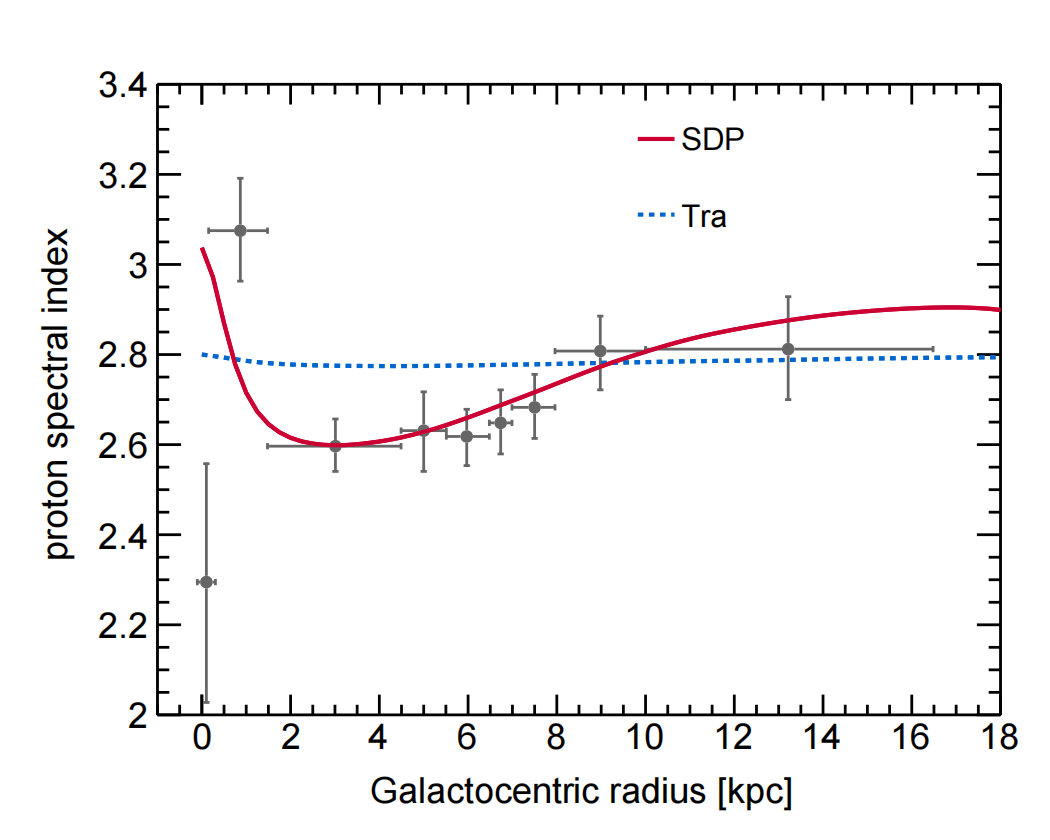}

\includegraphics[width=0.49\textwidth]{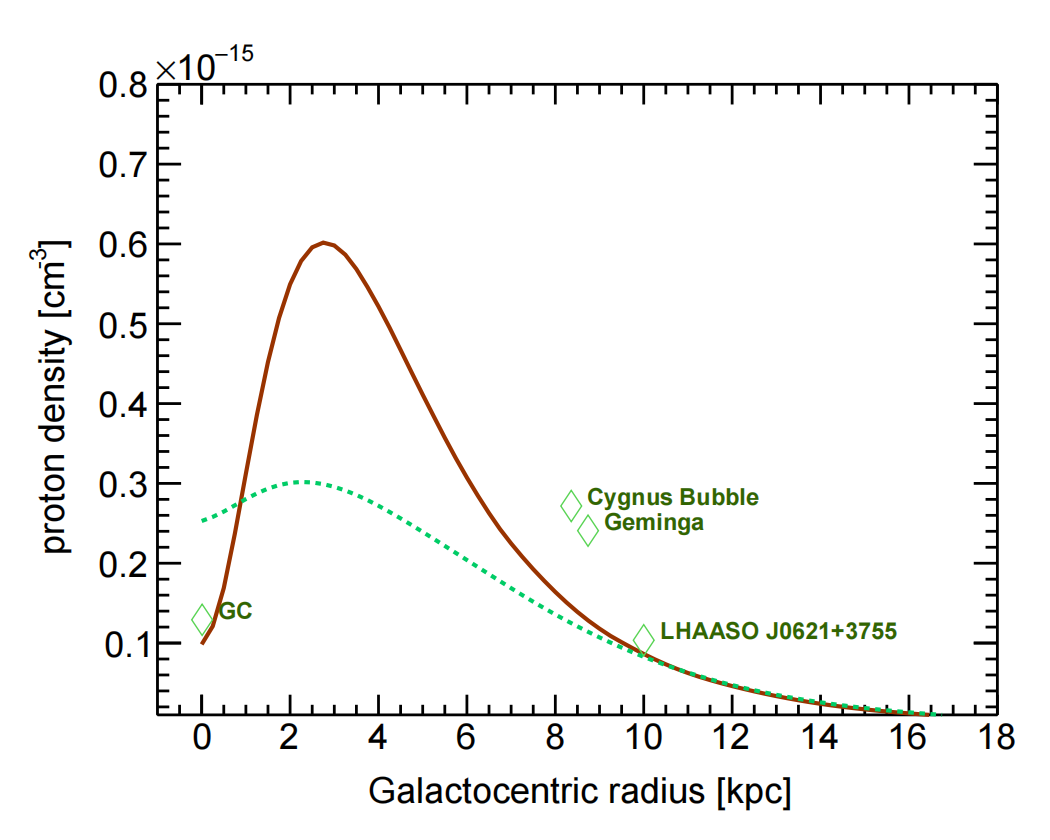}
\includegraphics[width=0.49\textwidth]{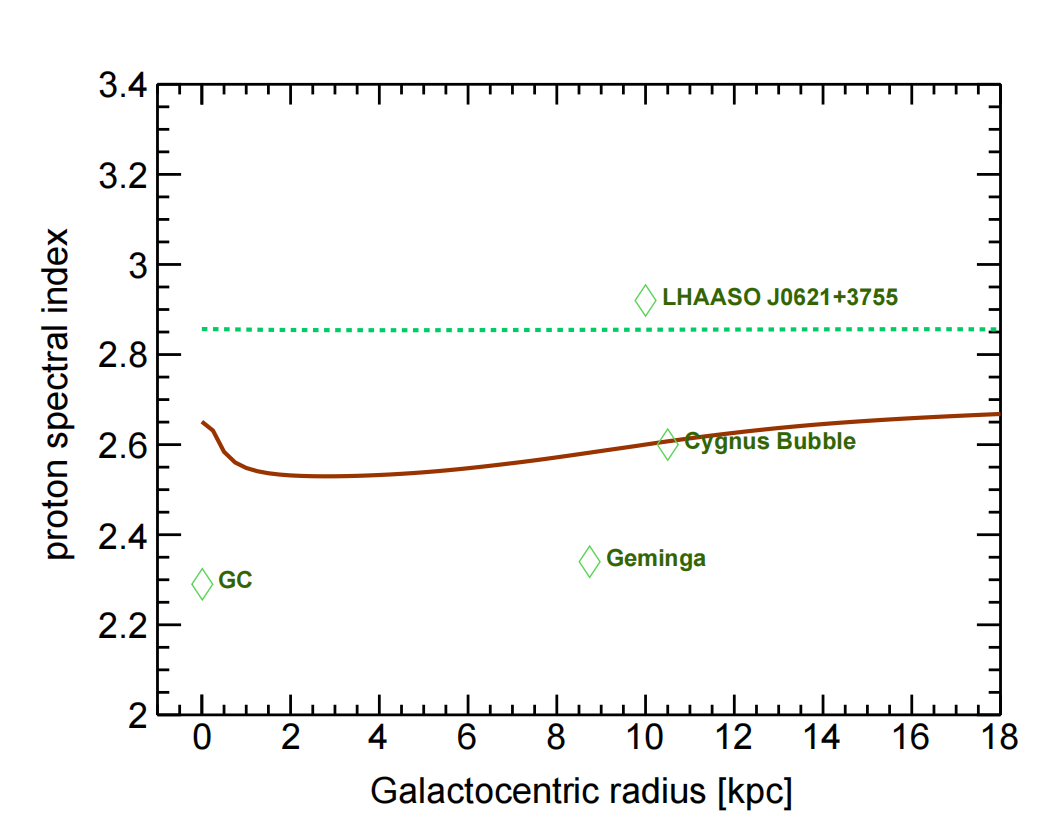}
\caption{Radial distribution across the Galaxy of proton flux integrated from 10 GeV to 150 GeV (upper left), and proton spectral index from 10 GeV to 150 GeV (upper right), compared with that inferred from Fermi-LAT $\gamma$-ray data \citep{2016ApJS..223...26A}. The solid lines represent results from the SDP model, while the dashed lines depict the traditional propagation model calculations. The lower two panels are proton density (lower left) and spectral index (lower right) at above 100 TV, with inferred values obtained from four observed TeV halos.
}
\label{fig:gamma}
\end{figure*}

\begin{figure*}[!ht]
\centering
\includegraphics[width=0.85\textwidth]{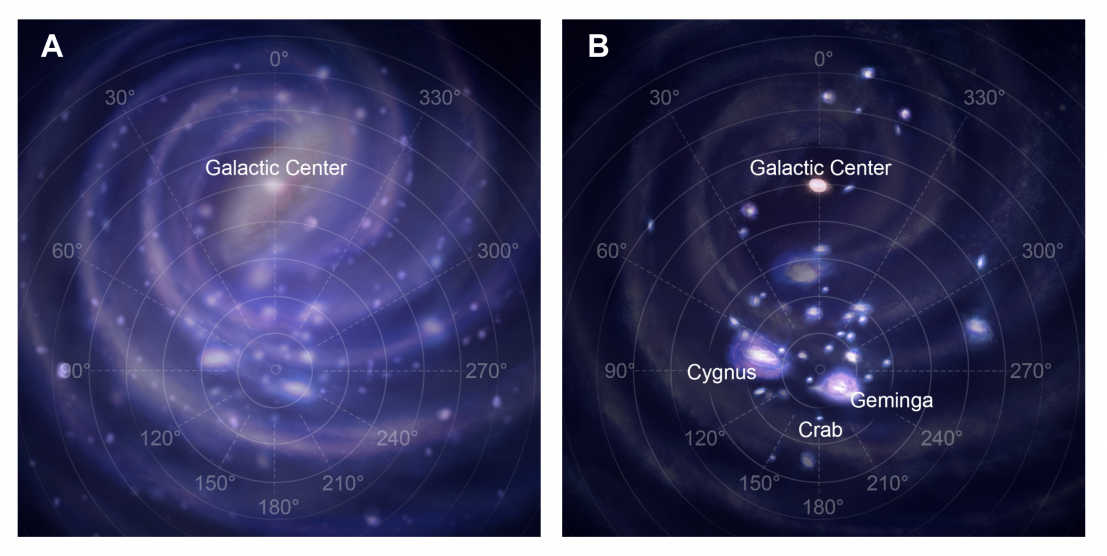}
\caption{The cartoon of diffuse $\gamma$-ray distribution for tens of GeV and TeV as left and right panel respectively.
}
\label{fig2:cartoon}
\end{figure*}

\begin{figure}[!htb]
\centering
\includegraphics[width=0.49\textwidth]{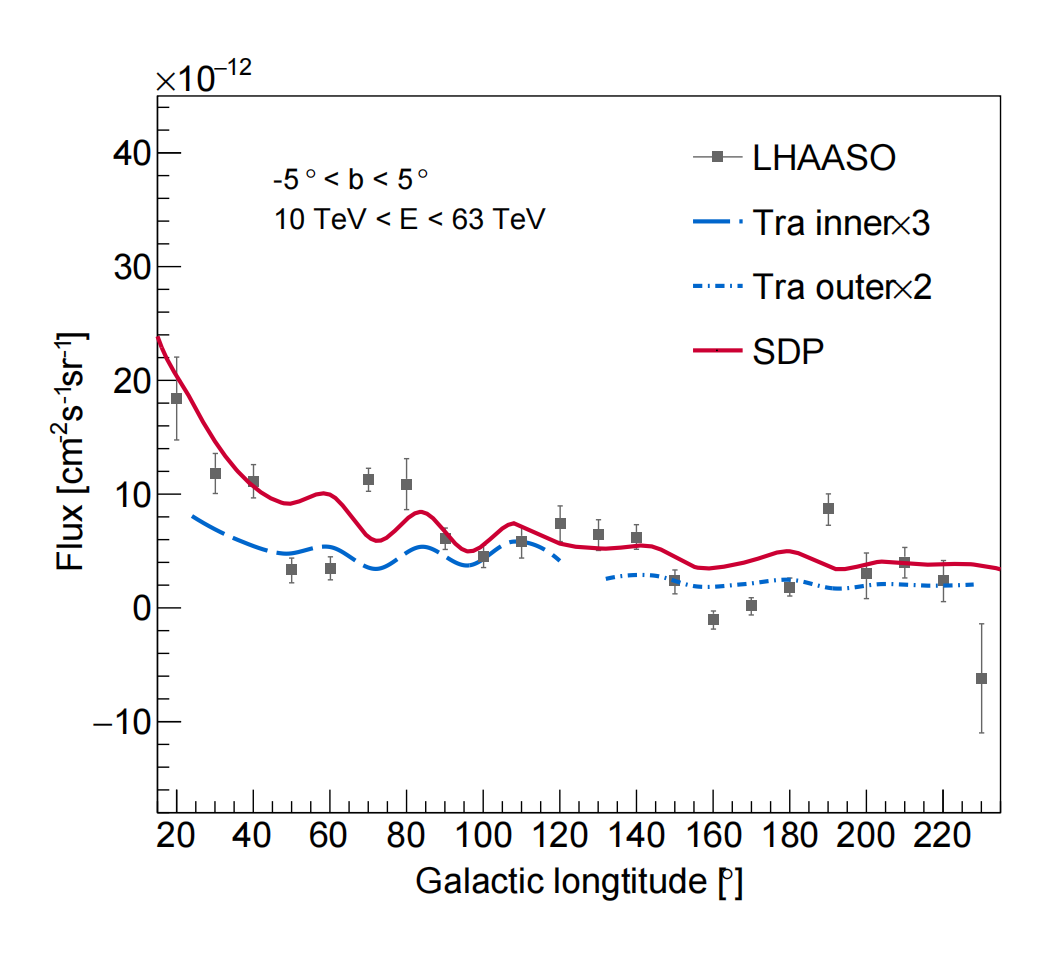}
\caption{
The flux of diffuse $\gamma$-ray emission in the regions of $|b|<5^{\circ}$, under the SDP and traditional propagation models, compared with data points from LHAASO measurement \citep{2023PhRvL.131o1001C}. To compare with the data, the fluxes calculated with the traditional propagation model have been multiplied by 3 and 2 for the inner and outer Galaxy regions, respectively.
}
\label{fig:diffuse}
\end{figure}

The charged-particle spectra within the solar system lend support to the physical interpretation illustrated in Figure \ref{fig:cartoon}. Following we extend our calculations to encompass the entire Galactic plane to investigate the spectral index and density variations of CRs. These variations have implications for the distribution of the diffuse gamma-ray emission throughout the Galaxy. In above calculations of charged particle spectra, we implemented a division of the source distribution centered at the Sun. Here we employed multiple centers spaced at intervals of 0.25~kpc along the Galactic plane. Each individual center allowed for the division of the source distribution $f(r,z)$ into two distinct zones: a spherical space with radii $ r_s <$ 0.6~kpc representing local nearby sources, and $ r_s >$ 0.6~kpc representing remote sources. By repeating the calculation using the same set of parameters as the solar observational calculations, we obtained the proton density and spectral indices at various Galactocentric distances.


The upper panel of Figure \ref{fig:gamma} illustrates the spatial distributions of proton densities and spectral indices within the energy range of 10 GeV to 150 GeV, along with the corresponding results from Fermi-LAT all-sky $\gamma$-ray data \citep{2016ApJS..223...26A}. It is evident that both the the proton density gradient and spectral indices predicted by the SDP model well match the observed steepness. Conversely, traditional transport models fail to account for spectral variations across the Galaxy due to their assumptions of uniform diffusion properties and a uniform injection spectrum from cosmic ray sources.

The lower two panels of Figure \ref{fig:gamma} display the proton density (lower left) and spectral index (lower right) above 100 TeV, along with inferred values obtained from several observed TeV halos. Notably, several well-established TeV halos, such as Geminga \citep{2017Sci...358..911A}, Cygnus Bubble \citep{2023arXiv231010100L}, Galactic Center \citep{2016Natur.531..476H} and LHAASO J0621+3755 \citep{2021PhRvL.126x1103A}, have been identified. We have included the high-energy spectral indices of these sources in their respective panels, and details about the TeV halos are listed in Table \ref{tab:halo}. Additionally, we have fine-tuned the diffuse coefficient of the SDP model to better match experimental measurements, enabling us to compute the proton density at the locations of these halos.

\begin{table*}[!htb]
\caption{Information regarding TeV halos. The first three columns indicate the location of each halo. The $D$ column represents the experimentally measured diffusion coefficient at the energy of $E_{p}$.  }\label{tab:halo}%
\begin{tabular*}{\textwidth}{@{\extracolsep\fill}ccccccc}
\hline
  & Gal Long  & Gal Lat  &  $r_{\odot}^\dagger$ &  $D$ &  $E_{p}$ &ref.  \\
  & [$^\circ$] & [$^\circ$]&  kpc&  $10^{28}cm^2/s$ & TeV  &    \\
\hline
 Galactic Center & 359.94  &  -0.04  & 8.5 & 60 & 10 &  \citep{2016Natur.531..476H}\\
 Geminga & 195.34 & 3.78 & 0.25 & 0.45 & 100 & \citep{2017Sci...358..911A}\\
LHAASO J0621+3755 & 175.76 & 10.95 & 1.6 & 0.89 & 160& \citep{2021PhRvL.126x1103A}\\
 Cygnus Bubble & 79.62 & 1.16 & 1.46 & 0.03 & 1 & \citep{2023arXiv231010100L}\\
 \hline
 $\dagger$ distance to the Sun.
\end{tabular*}
\end{table*}

This results further implies the presence of spatial dependence in the diffuse $\gamma$-ray spectrum at very high energy. To effectively illustrate this point, we provide a schematic diagram in Figure \ref{fig2:cartoon}, which depicts the nearly uniform and continuous nature of low-energy diffuse $\gamma$-ray emissions, while high-energy emissions are primarily influenced by local sources that may exhibit significant variations. The veracity of this prediction can be assessed through upcoming observations carried out by LHAASO, presenting a valuable opportunity to validate the precision of our calculations.

We present a novel scenario regarding the origin of CRs that significantly deviates from the traditional propagation model. We consider two distinct components, Comp-A and Comp-B, contributing to the CR fluxes. Comp-A originates from distant sources scattered across the outer diffusive zone, while Comp-B originates from local sources. In contrast, the traditional propagation model suggests the existence of only one component primarily originating from distant sources, as depicted in Figure \ref{fig:p_tra} in the appendix. Additionally, under traditional propagation models, diffuse $\gamma$-rays from GeV to PeV exhibit isotropic behavior throughout the Galaxy. However, our model predicts a different behavior. Figure \ref{fig:diffuse} showcases the model-calculated diffuse $\gamma$-ray flux in the 10 TeV - 63 TeV energy range. It is evident that the flux predicted by the traditional propagation model significantly underestimates the observations made by LHAASO. Conversely, the SDP model provides an average flux that aligns well with the experimental data. It is worth noting that experimental data on the flux intensity around $60^\circ-80^\circ$ exhibit higher values compared to the predictions of the SDP model. This discrepancy could be attributed to significant differences in the local source energy spectrum or truncation, as well as variations in the average source distribution.

Consequently, we emphasize that high-energy phenomena are predominantly influenced by local sources. Further measurements and observations in the future may help unveil the underlying the characteristics of source(s) behind these deviations.
Radio emission which is related with the electron and positron components of CRs should also be useful in testing the propagation of CRs \citep{haslam1981galactic,strong1978galactic,2011A&A...534A..54S,2018MNRAS.475.2724O}. While the uncertainties associated with the magnetic field models of the Milky Way may still be large, the multi-frequency radio data are found to be able to constrain some models (parameters) of the CR propagation \citep{2011A&A...534A..54S,2018MNRAS.475.2724O}. We leave the detailed multi-wavelength study of the diffuse emission from radio to ultra-high energy gamma rays under the SDP scenario in other work \citep{He2024}.

\section{Conclusions}
This study presents a novel CR propagation pictures to elucidate CR origin. In this scenario, the source halo plays a key role in the formation of Comp-A and Comp-B. Comp-A originates from remote sources and exhibits long-term, global, and universal characteristics, similar to the traditional propagation results. On the other hand, Comp-B originates from nearby sources and displays fresh, local, and spatially-dependent behavior within the galactic disk, which is lost in the traditional model.
Based on the SDP model, we validate this scenario by investigating the fluxes contributed from sources in the local region ($r_s <$ 0.6 kpc) and the distant region ($r_s >$ 0.6 kpc). Model calculations successfully reproduce the common characteristic of the two-component structure observed in all CR messengers, including primary, secondary, and their ratio spectra. Therefore, these calculations provide support for our proposed concept. Additionally, the diffuse $\gamma$-ray emissions observed throughout the entire Galaxy further bolster this physical framework. We anticipate that the diffuse $\gamma$-ray fluxes are spatially dependent due to the dominance of local sources in Comp-B. We strongly encourage the LHAASO/HAWC experiment to divide the entire sky into smaller segments and measure $ \gamma$-ray flux, which can serve as a means to test our predictions.

\begin{acknowledgments}
This work is supported by the National Natural Science Foundation of China (Nos. 12275279 and 12220101003), the China Postdoctoral Science Foundation (No. 2023M730423), and the Project for Young Scientists in Basic Research of Chinese Academy of Sciences (No. YSBR-061).
\end{acknowledgments}

\appendix

\section{Traditional Propagation Model}\label{secA1}
\setcounter{table}{0}
\setcounter{figure}{0}
\renewcommand{\thetable}{A\arabic{table}}
\renewcommand{\thefigure}{A\arabic{figure}}
The proton spectrum and B/C flux ratio under the traditional propagation model are depicted in Figures \ref{fig:p_tra}. The parameters of the traditional propagation model are listed in Table \ref{tab_tra:para}.


\begin{table*}[!htb]
\caption{Parameters of traditional propagation model}\label{tab_tra:para}%
\begin{tabular*}{\textwidth}{@{\extracolsep\fill}cccccc}
\hline
$ D_0  $  & $ \delta_{0}$  & $ N_m$  & n & $ v_{A}$ & $ z_0$ \\
 $ [cm^{-2}~s^{-1}]$ & &  & &  $ [km~s^{-1}]$ & [kpc] \\
 \hline
 $4 \times 10^{28}$ & 0.4  &  0.24  & 4.0 & 6 & 4.5 \\
\hline
\end{tabular*}
\end{table*}

\begin{figure*}[!htb]
\centering
\includegraphics[width=0.48\textwidth]{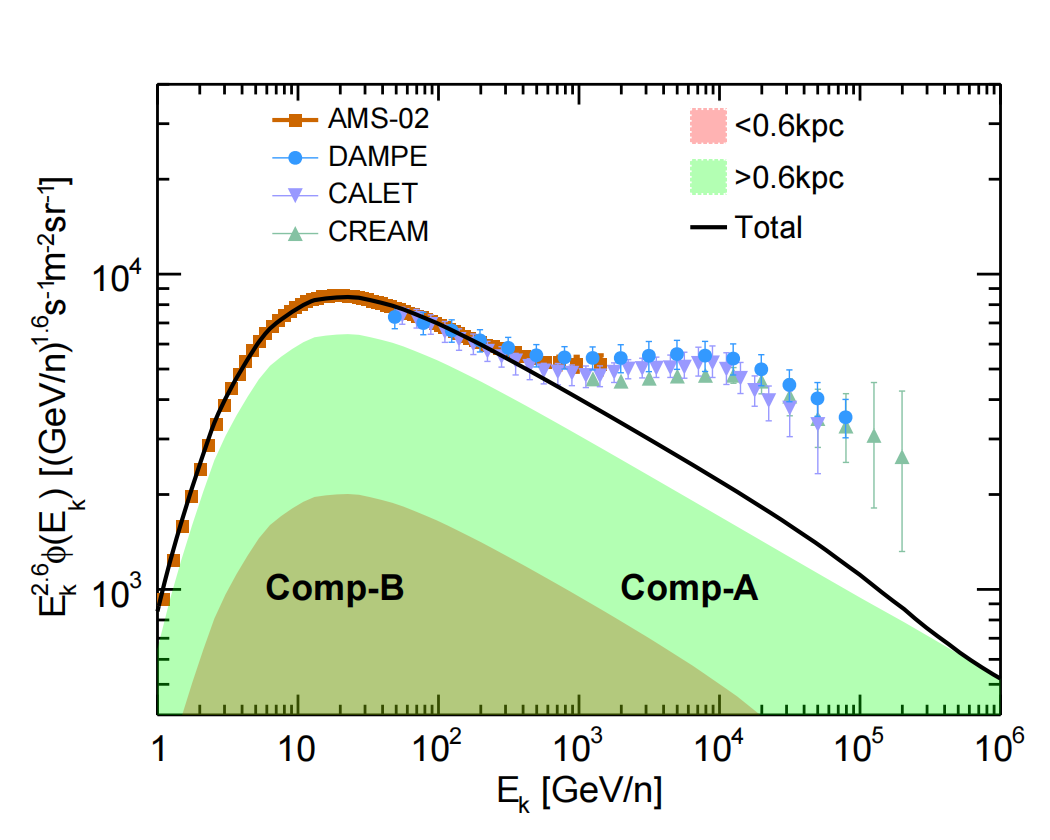}
\includegraphics[width=0.48\textwidth]{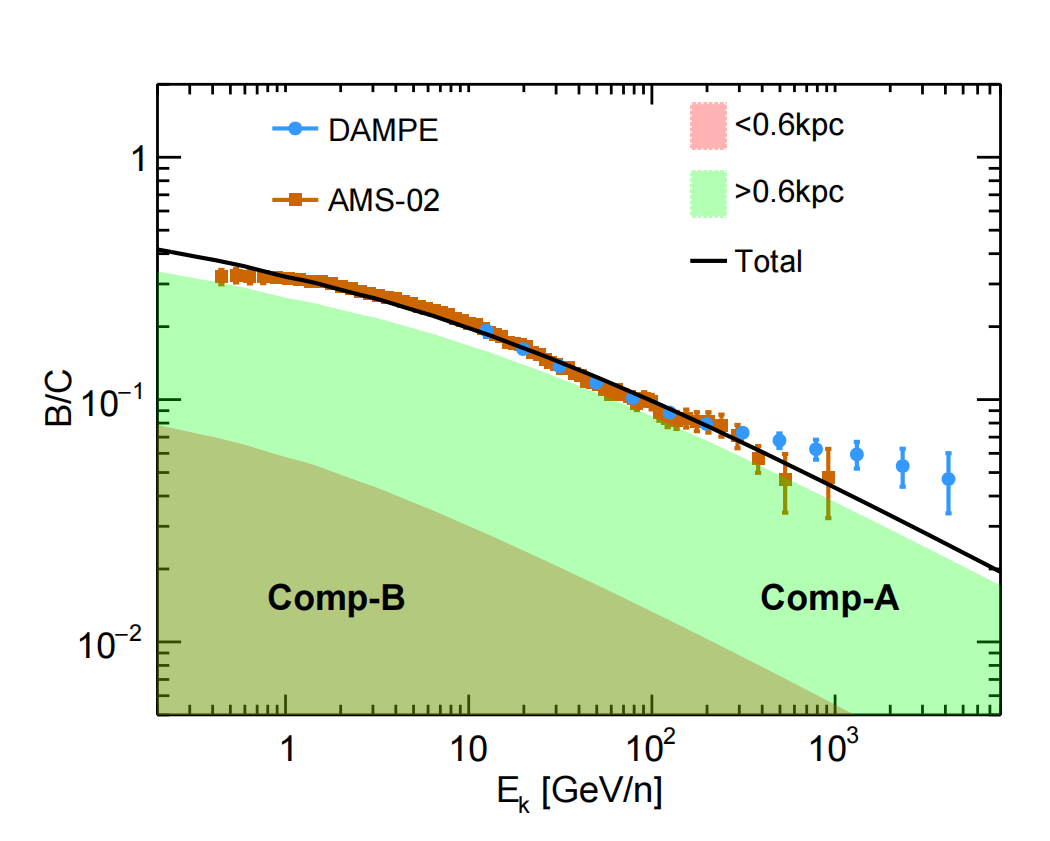}
\caption{
 Proton spectra and b/c flux ratio, under traditional propagation model. The propagation parameters are listed in Table \ref{tab_tra:para}. 
}
\label{fig:p_tra}
\end{figure*}

\nocite{*}
\bibliographystyle{unsrt_update}
\bibliography{apssamp}

\end{document}